# Anomalous microwave emission from spinning nano-diamonds around stars

**Authors:** J.S. Greaves[1]*, A.M.M. Scaife[2], D.T. Frayer[3], D.A. Green[4], B.S. Mason[5], A.M.S. Smith[6]

**Affiliations:**

[1]School of Physics & Astronomy, Cardiff University, 4 The Parade, Cardiff CF24 3AA, UK.

[2]Jodrell Bank Centre for Astrophysics, AlanTuring Building, Oxford Road, Manchester M13 9PL, UK.

[3]Green Bank Observatory, PO Box 2, Green Bank, WV 24944, USA.

[4]Astrophysics Group, Cavendish Laboratory, J.J. Thomson Avenue, Cambridge CB3 0HE, UK.

[5]National Radio Astronomy Observatory, 520 Edgemont Road, Charlottesville, VA 22903, USA.

[6]DLR, Institut für Planetenforschung, Rutherfordstraße 2, 12489 Berlin, Germany.

*Correspondence and material request to: GreavesJ1@cardiff.ac.uk.

**Several interstellar environments produce 'anomalous microwave emission' (AME), with a brightness-peak at tens of gigahertz[1]. The origins of AME are uncertain – rapidly-spinning nano-particles could emit electric-dipole radiation[2], but polycyclic aromatic hydrocarbons (PAHs) proposed as the carrier are now found not to correlate with Galactic AME[3,4]. The difficulty is to identify co-spatial sources over long lines of sight. Here we identify AME in three proto-planetary discs. These are the only known systems that host hydrogenated nano-diamonds[5], in contrast to very common detection of PAHs[6]. Spectroscopy locates the nano-diamonds close to the host-stars, at physically-constrained temperatures[7]. Developing disc models[8], we reproduce the AME with diamonds 0.75-1.1 nanometres in radius, holding ≤1-2% of the carbon budget. The AME:stellar-luminosity ratios are approximately constant, allowing nano-diamonds to be ubiquitous but emitting below detection thresholds in many star-systems. This can unify the findings with similar-sized diamonds found within solar system meteorites[9]. As nano-diamond spectral absorption is seen in interstellar sightlines[10], these particles are also a candidate for generating galaxy-scale[3] AME.**

Substantial discs comprising gases and dust-grains orbit around stars up to a few million years old, giving clues to the origins of the solar system and extra-solar planets. The discs evolve during planet-formation[11,12], with different properties observed for luminous stars such as Herbig A-type emission-line (HAe) objects, and 'classical T Tauri' stars (CTTS) akin to the young Sun. The solid content is mainly traced by thermal radiation from grains, with superposed spectral features from small particles, down to the size of PAH molecules.

One of the PAH spectral peaks is at 3.3 μm, and spectroscopy in this infrared band has also uncovered[13] a few sources with peaks at 3.43 and 3.53 μm. These features[5] are now identified with hydrogenated nano-diamonds[14], which are of special interest as a connection to solar system nano-diamonds in meteorites[9]. These particles may have formed under conditions of high pressure, shocks or vapour deposition[15], either internal or external to the proto-solar nebula. However, systematic searches[5,16] of over 80 HAe stars have shown only three hosting nano-diamonds, in contrast to PAHs found around many HAes[6] and some CTTS[17]. The diamonds could still be very common, but only rarely sufficiently excited to produce infrared spectra, perhaps by the most luminous host-stars[5].

In AME theory, the nano-particle carrier is not specified, and candidates have been extensively sought empirically. The imperfect correlation of PAH and AME sightlines[4] has led to new suggestions, including small hydrocarbons that can reproduce both AME and diffuse interstellar absorption bands[18] (DIBs), or silicate/iron nano-grains[4,19]. While supported by sophisticated models, these searches are hampered by not knowing if the spectral and AME sources are co-spatial. Here we address this by identifying AME in circumstellar discs, with accurate locations and well-measured physical properties. In particular, we find AME only in discs hosting hydrogenated nano-diamonds, where surface C-H bonds can provide suitable electric dipole moments. As many discs host PAHs, the nano-diamonds are favoured as the AME carrier.

Using a broad span of radio bands quasi-simultaneously helps to distinguish AME from other mechanisms (including time-variable processes). AME has a distinctively peaked spectrum in the microwave regime, while free-free electron-transitions in winds and jets, gyrosynchrotron flux from spots on stellar surfaces, and thermal emission from dust all follow power-laws. Two independent radio surveys were made here, covering 9 HAe systems observed at the Australian Compact Telescope Array (ATCA), and 5 systems with primaries[11] >1 $M_{Sun}$ observed at the 100m Robert C. Byrd Green Bank Telescope (GBT). The ATCA interferometer filters out any extended structure, with beams of ~35 down ~3 arcseconds at 5.5-97 GHz isolating the discs; some non-contemporaneous archival data were also processed to fill out frequency coverage. The GBT scans an ~25 arcsec beam at 26-40 GHz to filter out extended emission, producing photometry in four simultaneous bands. Follow-up measurements were made with the Arcminute Microkelvin Imager (AMI) at 16 GHz and GBT at 72 GHz; all new radio data are shown in Figures S1-S3.

Contributing signals from dust and winds were first subtracted, and then the AME residuals for three candidate discs were fitted. The subtraction procedure sought to minimise residuals over all wavelengths, within the constraint that these should not be negative (within the errorbars, which fold in noise, calibration, and temporal variability). The three sources remaining are thus those where AME is essential in order to reproduce the microwave emission. Full details of procedures are given in the Additional Information (Figures S1-S3 and Tables S1-S3).

The AME flux-profiles are shown in Figs 1-3. In the V892 Tau system, the AME peaks at a frequency around 25 GHz, and has a maximum amplitude of ≈1 mJy (over half the total signal, Fig. S3). The GBT residuals are independently confirmed from ATCA data[20]. HD 97048 has a similar AME amplitude, inferred from our two ATCA points, while the peak frequency is shifted slightly lower, to ≈20 GHz. In the luminous MWC 297 system, the peak frequency lies at ~50 GHz (between our ATCA and GBT bands), and the AME amplitude is much higher, at ~30 mJy.

**Fig. 1**. *(Left panel:)* Data points for V892 Tau, with dust and wind model subtracted to leave the AME residual. Error bars include the range of allowed subtracted wind+dust signals, plus flux uncertainties (or a minimum of 10% calibration uncertainty), added in quadrature. Upper limits are +2σ (gaussian statistics are used throughout). Dashed line shows the Maximum Likelihood (ML) fit; the solid line shows the model using the parameters' expectation values; thin lines are 24 samples randomly drawn from the posterior distribution. *(Right panel:)* Parameter space for model variables *a* (nano-diamond radius) and nano-diamond:carbon abundance, showing marginalised posterior probability. Solid lines mark positions of ML parameter values; dashed lines are 16, 50, 84% quantiles on the 1D posteriors (equivalent to -1σ, mean, +1σ); contours show 68, 95, 99% levels on the 2D posterior.

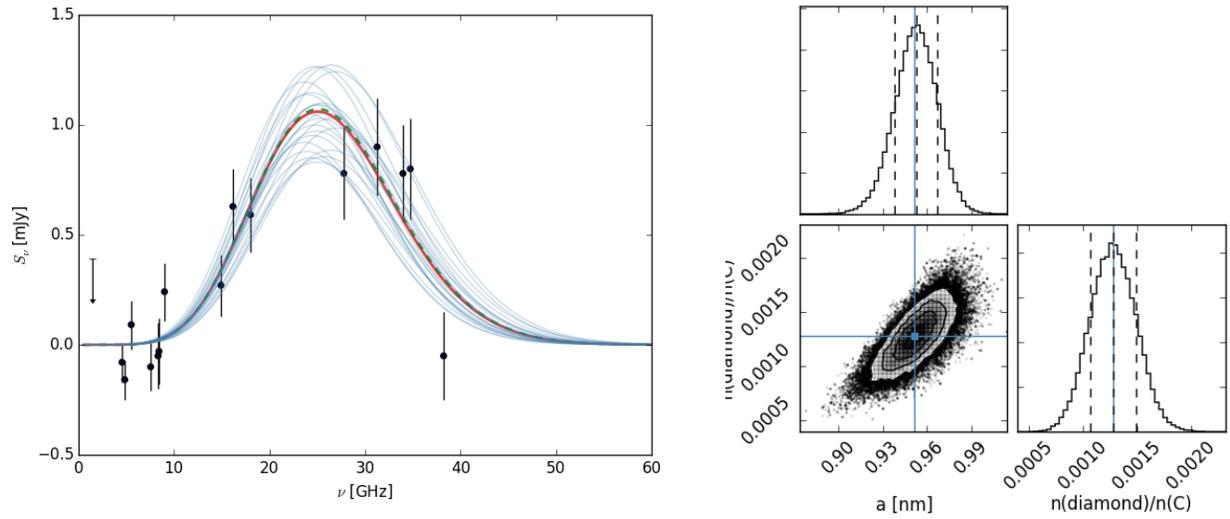

**Fig 2.** AME data points and models for HD 97048; details as in Fig. 1.

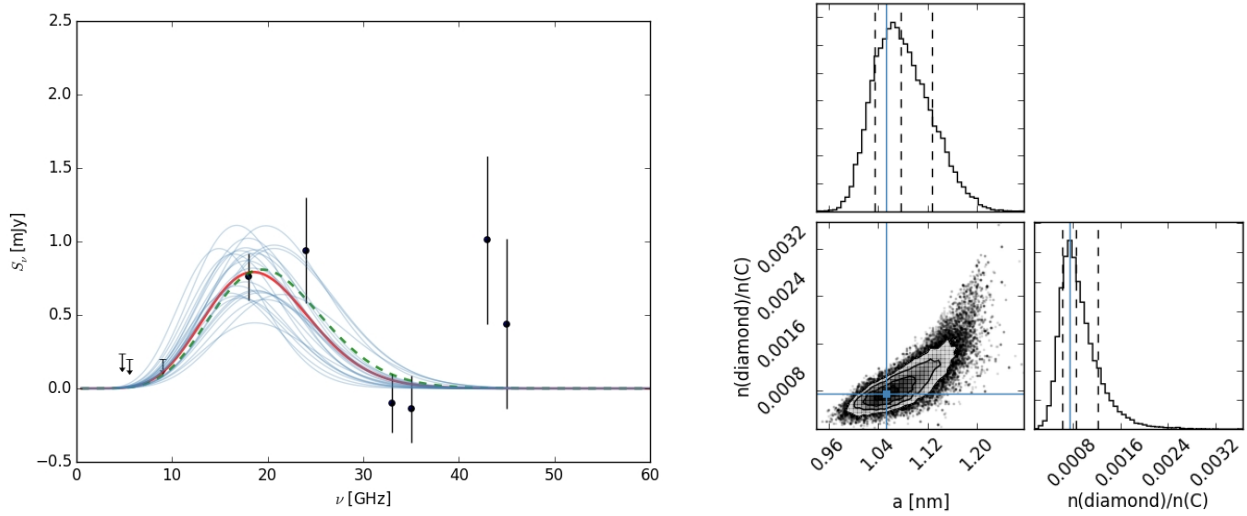

**Fig 3.** AME data points and models for MWC 297; details as in Fig. 1.

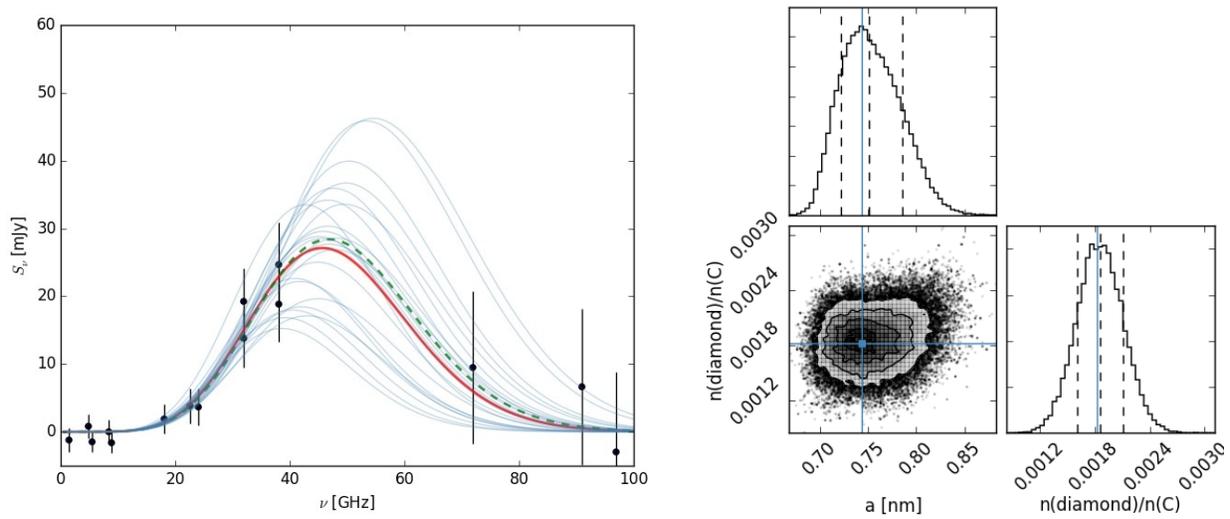

Model AME spectra were constructed from a disc formulism[8] for spinning nano-particles. For a sphere large enough to behave classically, the power radiated is

$$P(a,\omega) = (4/9)\, d^2(a)\, \omega^4 / c^3 \qquad (1)$$

for particle radius $a$, spin frequency $\omega$ and electric dipole moment $d$. Prior work[2,8] set $d = N^{1/2} d_0$ for N randomly-oriented surface dipoles, with e.g. a C-H bond having $d_0 \approx 0.4$ Debye. However, in symmetric diamondoids these dipoles would self-cancel, so here $d$ is assumed dominated by ionised and asymmetric forms[21,22] with $d \sim 0.5$-$5.7$ Debye. In Table 1, we fit with a mid-range $d = 1.5$ Debye, and subsequently scale radiated power $P$ by $d^2$ (Eq. 1).

For a Boltzmann distribution of spin rates given by

$$f(\omega) = 4\pi\, (3/2\pi)^{3/2}\, (\omega^2 / \langle\omega^2\rangle^{3/2})\, \exp(-3/2\, \omega^2 / \langle\omega^2\rangle), \qquad (2)$$

the expectation value of the emission frequency (in radians/sec) for temperature $T$ and moment of inertia $I$ is

$$\langle\omega^2\rangle^{1/2} = (3 k_B T / I)^{1/2} \qquad (3)$$

which varies as $T^{1/2} \rho^{-1/2} a^{-5/2}$; a mass-density $\rho \approx 2.5$ g/cm$^3$ is adopted around C-atoms spaced[23] by 0.2 nm. Model spheres were assumed, to approximate to diamond chunks within meteorites and to match prior work[8], but noting that $d \neq 0$ actually requires asymmetry or charge. We fit here for one characteristic size $a$ rather than a distribution, as cage-structured diamonds have quantised sizes[23].

The AME spectrum is then given by

$$L(a) = n(\text{diamonds})/n(\text{H}) \int 2\pi R\, \Sigma(R)/m_H\, dR \,.\, P(a,\omega) \,.\, 2\pi f_\omega(\omega,R), \qquad (4)$$

integrated over a disc with radial mass-surface-density $\Sigma(R)$. Two of the nano-diamond surface-profiles are unresolved[24,25], with V892 Tau fits[7] declining as an $R^{\sim -0.3}$ function (but this is subject to assumed particle size). Table 1 results are for a flat $\Sigma(R)$, while steeper profiles tested for V892 Tau yielded lower nano-diamond abundances.

Particle temperatures were taken to be in local thermodynamic equilibrium, although this is not strictly valid for sizes under 2 nm with transient heating[26]. Hydrogenated nano-diamonds are only stable[7] at T ≈ 800-1400 K, so we set the observed $R_{inner}$ (V892 Tau) or $R_{outer}$ (HD 97048, MWC 297) to correspond to the upper or lower temperature bound, and fixed the other radius via the equilibrium condition of T proportional to $R^{-0.5}$. The warm temperatures probably require the nano-particles to float in the largely atomic layers above and below the disc mid-plane.

The AME fits (Figs 1-3, Table 1) are functions of only two free parameters, the nano-diamond size and the abundance (against total carbon, with cosmic C:H = $4 \times 10^{-4}$). The model diamond spheres are found to have similar radii, of 0.75-1.1 nm, in all three systems. The different AME peak-frequencies have emerged naturally from the varying disc sizes and stellar luminosities, without requiring diverse nano-particles. These model spheres equate to ~200-700 C-atoms, while for equal counts, the smallest tetrahedral equivalent would be a 6-layer pyramid[23], 1.8 nm on a side, and 'boxy' forms[21] would be longer. In the case of HD 97048, the nano-diamond spectra (Fig. S4) have a possible fit with a comparable 5-layer pyramid[23]. Similarly-scaled nano-diamonds within meteorites[9] are often around 2-3 nm across. One uncertainty in our model is the excitation temperature of the particles, which could be much lower than the kinetic temperature in diffuse gas[18], as for example in high layers of the disc. We implicitly took $T$ above to be the kinetic temperature, but in an extreme alternative where the particles were excited to only ~10 K (as estimated for diffuse interstellar gas[18]), their sizes would shrink from 0.75-1.1 nm to ~0.3-0.45 nm (equation 3). The lower size bound corresponds to the smallest possible nano-diamond species, adamantane ($C_{10}H_{16}$), which may contribute[23] to the infrared spectrum of HD 97048.

**Table 1.** Stellar and disc parameters and AME model results. ML and Ex indicate maximum likelihood and expectation fits. These solutions are for a flat surface-density profile, with disc-mass estimates based uniformly on dust fluxes measured near 110 GHz (optically-thin regime); the abundances (diamond:carbon fraction) scale inversely with $M_{disc}$. The solutions also adopt a fixed dipole moment of 1.5 Debye; the abundance scales as $d^{-2}$ (Eq. 1), with different diamond forms having $d \approx 0.4$-5.7 Debye.

|  | V892 Tau | HD 97048 | MWC 297 |
|---|---|---|---|
| $d$, star distance (parsecs) | 142 ± 14 | 158 ± 16 | 250 ± 50 |
| $T_{eff}$, effective temperature (kelvin) | 11,200 | 15,000 | 24,500 |
| $L_*$ (solar luminosities) | ~80 (40-96) | ~40 (30-50) | ~21,000 (12,000-32,000) |
| $M_*$ (solar masses) | ~5.5 (~8 AU binary) | 2.5 | ~10 |
| $M_{disc}$ (solar masses) | 0.035 | 0.11 | 0.4 |
| $R$(diamonds) in model (AU) | 10-30 | 5-15 | 40-120 |
| $a$, size of model diamonds (nm) | 0.95 (ML)<br>0.952 ± 0.015 (Ex) | 1.05 (ML)<br>1.07 ± 0.05 (Ex) | 0.74 (ML)<br>0.75 ± 0.03 (Ex) |
| diamond:carbon fraction (%) | 0.13 (ML)<br>0.13 ± 0.02 (Ex) | 0.07 (ML)<br>0.08 ± 0.03 (Ex) | 0.18 (ML)<br>0.19 ± 0.03 (Ex) |
| $L_{AME}$ (solar luminosities) | 1 x $10^{-7}$ | 1 x $10^{-7}$ | 2 x $10^{-5}$ |

The nano-diamond abundances (Table 1) would scale up to at most 1-2% of the total carbon, for dipole moments reduced to as small as 0.4-0.5 Debye (single C-H bond or single charge). These abundances are less than or equal to the ~1-3% diamond:C ratios that reproduce the interstellar 3.47 μm nano-diamond band, of similar absorbance[21] to the 3.43/3.53 μm features. If the total budget for carbon locked up in small species is interstellar-like[8], at ~5%, the nano-diamonds in the discs also fit well inside this constraint.

Our results closely associate AME from circumstellar discs with the presence of nano-diamonds. The total number of discs we surveyed with GBT/ATCA was 39, of which 14 have luminous host-stars (Tables S1-S3). AME was found *only* in the three discs with nano-diamond signatures, not generically in the luminous-star discs (of which 85-100% exhibit PAHs, Fig. S5) or in the wider star-sample. To estimate a false-alarm probability, we used the occurrence of nano-diamonds in comprehensive spectral surveys[5,16] of HAe discs, yielding 3/82, or a probability $P_{nano}$ = 0.037 per star. The probability in a 14-star sample of, by chance, picking the three stars with AME and finding they have nano-diamonds, and also finding that the remaining 11 stars do not, is then $P_{chance} = P_{nano}^3 (1 - P_{nano})^{11}$, or 0.003%. This is robust against removing individual

systems – e.g. CD-42 11721 is now known to be unusually distant (Table S3), so a better estimate would be $P_{nano}^3$ (1 - $P_{nano})^{10}$, yielding $P_{chance\ of}$ 0.0035%. More conservatively, we also checked biases towards discs with brighter dust and PAHs in our GBT/ATCA target selection, which could have eliminated some of the nano-diamond-search population. This could raise $P_{nano}$ to ≈0.06, and hence $P_{chance}$ to ≈0.01%, which is still very low.

In proto-planetary discs, the carrier of AME is thus strongly indicated to be hydrogenated nano-diamonds. PAHs appear much less probable, as our sample has numerous PAH-hosting discs that do not exhibit AME. In particular, there are discs brighter in PAH features (Fig. S6) than V892 Tau and HD 97048, and these discs do not show AME that could be expected if PAHs were the carrier. In contrast, there is a divide in parameter space between the nano-diamond hosts with AME, and the systems without either phenomena (Fig. S6).

We hypothesize that more distant and/or lower-luminosity star-systems are less detectable in the microwave regime, but still could host nano-diamonds at similar abundances – this would also help to link our findings with the diamonds found around the cool, lower-mass Sun. The ratios of AME to stellar luminosity are ≈(1-3) x $10^{-9}$ (constant within uncertainties, Table 1), even though the three stars span a range of ~500 in $L_*$. Hence it appears that the detection of AME is likely to be flux-limited in our sample (as shown Fig. S6, where only AB Aur modestly challenges the constant L(AME)/$L_*$ scenario at an upper limit of ≈0.3 x $10^{-9}$). There is no corresponding simple relationship of the nano-diamond spectral fluxes to host-star properties, after nearly four decades of study[5,13]. However, the three detected systems are among the four hottest stars in our sample (Table S3; CD-42 11721 is similarly hot but prohibitively distant). The three AME host-stars will thus be strong ultra-violet sources, the wavelength regime in which excitation bands for nano-diamonds[21,22] lie. For example, in the ~200 nm band[22], model stellar-surface fluxes[27] are strongly dependent on temperature, with an increase of only 10% in $T_{eff}$ raising this flux by a factor of 2. Hence a temperature threshold may prohibit detecting the IR spectral features of nano-diamonds around slightly cooler stars. Applying a cut-off of Teff ≈ 10,000 K for excitation of the infrared spectra, and also predicting microwave fluxes on the basis of source distance and the observed range of L(AME)/$L_*$, we find that no other systems in our sample that should show combined infrared spectra and AME. As excitation also occurs[21] near the ionization limit of ~8 eV, if charged nano-diamonds are the AME carriers, there may also be a link to detections of AME in environments with ionised gas[28].

Many other carriers of AME have been proposed[4,18,19], but discs are the only environment where the candidate particles have now been precisely located via spectroscopy. We do not rule out that other hydrocarbons can also be AME carriers, but note that nano-diamonds are widespread[23]. One model[18] proposes that small AME-carriers of only ~8-15 C-atoms could also reproduce diffuse interstellar bands (DIBs), and the smallest nano-diamond form (adamantane, $C_{10}H_{16}$) is in this regime. Further, the fullerene ion $C_{60}^+$ is the only identified[30] DIB carrier, and carbon 'onion' structures such as $C_{60}$ have been proposed as sites for nano-diamond formation[9]. $C_{60}$ is now known in one HAe disc, and also in two evolved-star envelopes that host nano-diamonds[29]. Hence, if fullerenes and similar species are rather ubiquitous and can provide viable production sites, this may be a route to generating nano-diamonds across the size-range of ~10-700 C-atoms that could explain AME from diffuse interstellar gas and from dense circumstellar discs.

Solar system nano-diamonds may have been made in the proto-solar disc and/or inherited from previous generations of evolved stars. In star-systems generally, disc evaporation and stellar winds could expel nano-diamonds back into the interstellar medium. This provides a testable hypothesis, where the 3.47 μm nano-diamond absorption features in dense interstellar clouds[10,23] may correlate with AME sightlines. Given a widespread distribution, nano-diamonds could thus present an alternate solution for the problem of poor correlation of AME and PAH distributions[4].

**Acknowledgments:** The NRAO is a facility of the National Science Foundation operated under cooperative agreement by Associated Universities, Inc. Infrared spectra are presented from the processed data archives of ESA's Infrared Space Observatory and NASA's Spitzer Space Telescope. A. Avison at JBCA reduced the ALMA observations of HD 97048. AMS gratefully acknowledges support from the European Research Council under grant ERC-2012-StG-307215 LODESTONE. We thank the staff of the Lord's Bridge Observatory for their assistance in the operation of the AMI. The AMI is supported by the University of Cambridge and the STFC.

**Author contributions:** JSG led the project, analysed GBT and ISO data, coded initial models, and drafted the paper. AMMS analysed ATCA data, contributed AME and coding expertise, and wrote modeling sections of the paper. DTF, DAG, BSM and AMSS contributed instrument, observation and software support and commented on the paper.

**Additional information**

Supplementary information is available for this paper.

Correspondence and requests for materials should be addressed to JSG.

**Competing interests**

The authors declare no competing financial interests.

*(Greaves: Anomalous microwave emission from nano-diamonds:)* **Additional information**

Radio Observations

The Australia Telescope Compact Array was used to observe 9 discs of HAe stars. The array was in the H214 hybrid configuration, with baselines in the range from 92 to 4383 metres between the six 22m antennas. Signals from the C/X and mm receivers were correlated using the Compact Array Broadband Backend. The survey was designed to search for AME, and observations used bands at 5.5, 8.8, 18, 24, 32, 38, 91 and 97 GHz, with bandwidths of 2048 MHz. In total, 28 datasets were obtained for the nine sources over 6-11 October 2010. For MWC 297, observations were made in all 8 frequency bands, each lasting from 6 to 130 minutes. Only four frequency bands were observed for HD 97048, for durations of 20-160 minutes. Multiple scans were made in all cases (except the two lowest frequencies observed for MWC 297). Uranus was used as the primary flux calibrator along with secondary calibrators (1934-638 for MWC 297 and 0537-441 for HD 97048; the latter is slightly variable in flux, so pairs of data points are shown for primary and secondary calibration). The data are publicly available under project code C2426. Archival data for HD 97048 were also processed to provide additional frequencies (from project code C1794, using a range of ATCA configurations).

Observations of V892 Tau were made at the 100m Green Bank Telescope in West Virginia,. Thirty HAe and CTTS systems were observed in a flux-limited survey covering discs with 1.3 mm flux >90 mJy in Taurus and Ophiuchus. The Ka-band receiver was used with the Caltech Continuum Backend; the CCB uses optimized detector circuits and 4 kHz beam-switching to suppress instrumental gain fluctuations. Four frequency channels are obtained at 26-29.5, 29.5-33.0, 33.0-36.5 and 36.5-40 GHz. Photometry utilised an on-the-fly nod, with four 10-second phases in a 70-second observation. Seven repeats of this sequence were made for V892 Tau on 21 April 2007, immediately after a skydip and calibration check, and before an observation of DL Tau. The latter showed a normal power-law spectrum across the 4 sub-bands with an index of 1.7 (in the convention of positive index for rising flux at higher frequencies), with correlation coefficient of r=0.99. Flux densities were established by observing the primary calibrators 3C 48 and 3C 147, which have power-law spectra (measured indices of -1.18 and -1.09 respectively). Neither calibrator was undergoing flux changes at similar frequencies at the time[31]. The individual data points for V892 Tau are shown in Fig. S1. The archived data are available under project code AGBT07A-038.

V892 Tau was also observed with AMI-LA, the Large Array of the Arcminute Microkelvin Imager[32]. This comprises eight 13m antennas sited at the Mullard Radio Astronomy Observatory at Lord's Bridge, Cambridge, UK. The telescope observes in the band 13.5–17.9 GHz with eight 0.75 GHz bandwidth channels, but the two lowest frequency channels have lower response in this frequency range and suffer from interference; the effective frequency here was 16.1 GHz. In total ten HAe/CTTS sources were observed in this band from July and September 2011. AMI-LA flux calibration is performed using observations of 3C286, 3C48 and 3C147, with I+Q flux densities for these sources in the AMI-LA channels consistent with the updated VLA calibration scale; polarization and airmass are also corrected for. Tests show fluxes are accurate to ≤5%.

MWC 297 was also observed with the GBT W-band receiver, on 26 September 2016 (project code AGBT16B-390). Fast scans of the telescope's 10-arcsecond beam were made to extract the source signal from the background level. The effective on-source time per pass was ~5 seconds, with 64 scans made in total. Sky conditions were good at 72 GHz, with zenith opacity of 0.26, but the beam size varied with telescope temperature; the source is point-like within this limitation (Fig. S2). The signal-to-noise ratio is 14, and the 72 GHz flux is 85 ± 11 mJy (for an error budget of 8% noise and 10% in calibration); the flux calibrator was 1751+0939.

All new results for the AME sources are presented in Table S1. Literature flux densities[20,33-40] from VLA, ATCA, CARMA, BIMA and ALMA were included in our analysis for consistency checks, to fill in frequency coverage, and to fit combined signals from wind plus dust (Fig. S3). We independently reduced an archival dataset[39] of HD 97048 from the Atacama Large Millimetre Array to generate a flux error at 106 GHz. For HD 97048 (CU Cha), the archival ATCA data (Table S1) were recently published[40]. Our results agree within the errors, but are systematically slightly higher in flux, probably due to our use of phase self-calibration of the field.

Radio Data Analysis

The total radio-frequency range in the analysis spans 1.4-115 GHz. The fluxes of dust and wind components in each system were fitted and subtracted to yield a *minimum* AME residual signal. The winds were characterised at frequencies below ~10 GHz, and temporal variability was included where possible (Fig. S3). The wind indices resulting from the fits are all within known bounds, which extend from -0.1 up to 0.6 in simple optically-thin geometries, increasing to 2 in optically-thick cases. Dust signals were fitted at the high frequency end, with maximised numbers of frequencies set to have no AME. In addition, dust spectral indices were confined to the range 2-4, appropriate for large to small grain sizes, emitting as blackbodies and inefficient greybodies respectively. An overall fitting requirement was that residuals after dust-plus-wind subtraction should not be negative, within errors. Although SR 21 is a weak AME candidate, this source is a pair, with the low 34 GHz flux[20] including only the A-component. Hence the GBT flux at 31.25 GHz (detected at only 3.5σ), suggesting a small candidate residual, appears to be mainly from the less-studied B-component (which has no independent IR spectroscopy).

The highest frequency for each source was also used to estimate masses of dust. The 106-115 GHz fluxes were scaled by source distance, and then converted to mass via a V892 Tau model[33]. Thus masses for HD 97048 and MWC 297 ignore any variations between host stars and disc geometries, but benefit from using the most optically-thin (longest) wavelength. Literature fluxes at 230-345 GHz lie below extrapolations to our dust fits by factors of 2±0.5, indicating similar optical depths in each case, and so reasonably robust results from the scaling approach.

Infrared Spectra

The heritage archive of the *Infrared Space Observatory (ISO)* yielded nano-diamond spectra of V892 Tau, MWC 298 and HD 97048 (Fig. S4). The 3.43, 3.53 μm features lie in the '1D' sub-band of the Short-Wave Spectrometer, and highly-processed data products[41], HDDP, are shown, where artificial fringing has been reduced. These spectra are discussed in the literature[42] and also have ground-based equivalents[5]. Fig. S5 shows additional archival spectra, from *ISO* and the *Spitzer* Space Telescope, to demonstrate PAH features around 6.2 μm observed for our sample of HAe stars. Another nano-diamond peak is predicted[23] at ~6.88 μm and its presence was suggested[42] for HD 97048, but in discs it can be blended with a 6.85 μm line of water vapour.

Table S3 lists fluxes and limits for the 3.53 μm nano-diamond feature for all the sources in our sample. The values listed are mainly from a survey[16] made with *ISO;* these agree only within ~60% with ground-based values[5] due to difficulties in absolute calibration, and so the errors were set conservatively at a 5σ level[16]. For T Tau, we estimated a limit from an archival *ISO* HDDP, and for SR 21 and IC 2087 IR we used the rms in ground-based spectral observations[45,46] to generate a similar error. Fig. S6 plots the AME detections and limits against the fluxes for the 3.53 µm nano-diamond and 6.2 µm PAH features (Table S3).

**Fig. S1:** data stream for V892 Tau observed with the GBT. The raw signal is plotted against observation sequence number, in the four sub-bands of the 26-40 GHz receiver. Errors on the individual data points range from 0.2-1.0 mJy, with highest uncertainty at higher frequencies where the sky is more opaque. The co-added signals in each band have errors of 0.15-0.17 mJy (Table S1), with different methods of weighting the points affecting these means by ≤0.05 mJy.

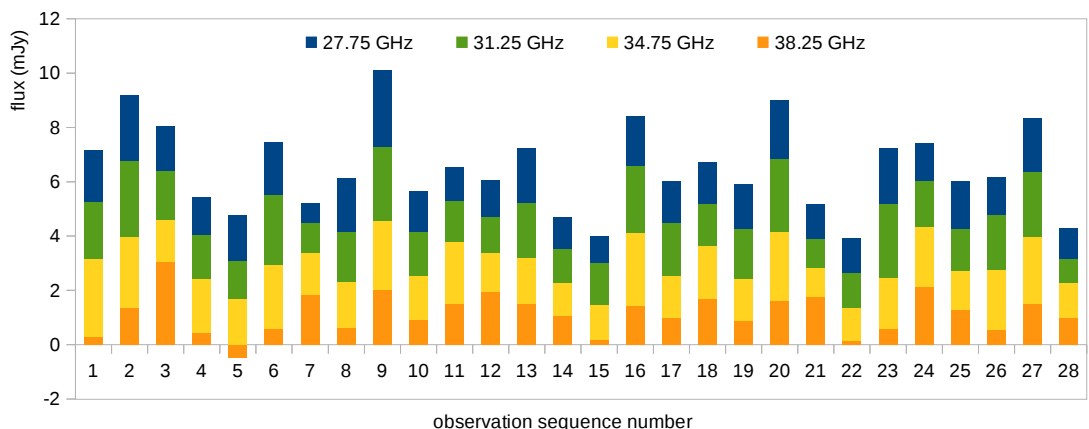

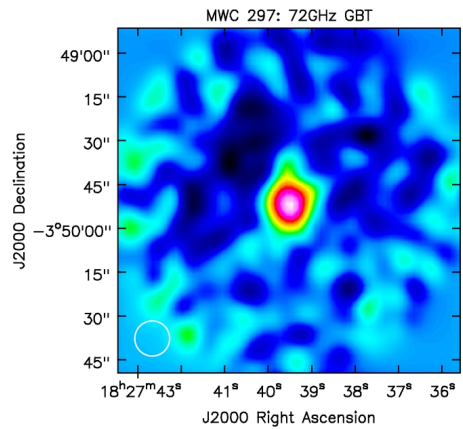

**Fig. S2.** GBT image at 72 GHz of MWC 297. The pixel sizes are 2 arcsec and the beam size is shown by the circle. The time-stream data were filtered to remove sky and instrument effects, re-gridded into RA,Dec., and smoothed for clarity with an 8 arcsec wide Gaussian profile. The noise increases on the edge of the map and as a function of radius within the map due to the daisy-scanning mapping pattern. The radii of the areas sampled in the scan pattern were 0.8-1.0 arcmin.

Statistics

The probability $P_{nano}$ was corrected for biases. ATCA targets were chosen to have bright PAHs, and e.g. the faintest 6.2 µm line in this sample has a flux of 6 $10^{-15}$ W/m$^2$; ~20-40% of the HAe stars in the *ISO* nano-diamond survey[16] fall below this. GBT targets were chosen to have 1.3 mm dust flux >90 mJy; ~50% of the *ISO* survey fall below this, e.g. using a mean spectral index ~2.5 to extrapolate 0.85 mm fluxes[48]. Weighting by the 9 ATCA and 5 GBT targets, the proportion of HAes that were searched for nano-diamonds but not for AME is ~30-45%, so $P_{nano}$ is de-biased from 0.037 to ~0.06. The probability of getting 3 nano-diamond hosts in a random sample of 14 HAe stars is then $P = P_{nano}^3 (1 - P_{nano})^{11} C_{14,3}$ where $C_{n,k} = n!/(n-k)!k!$ is the number of ways of drawing 3 stars out of 14. This is then divided by $C_{14,3}$, the number of ways of picking 3 AME-hosts out of 14 radio targets, to yield $P_{chance}$ for the 3 AME hosts to also have nano-diamonds. One multiple-source (Fig. 3e below) could be biased against AME detection, but omitting this object would only increase false-alarm probabilities by a factor $1/(1 - P_{nano})$, ≈ 1.04-1.06.

**Fig. S3.** Total flux (on a log scale) against frequency for each AME system. Archival and literature data are shown with small circle symbols, with large circles showing the new flux results (Table S1). Upper limits (2σ) are shown by triangles. Error bars are 1σ, and are set to minimum levels of 10% of measured flux, to account for uncertainties in calibration scales. The green dotted curves show the fits for combined wind and dust emission.

*(Top:)* V892 Tau. The dust component is fitted as a $\nu^4$ power-law spectrum scaled to 22 mJy at 100 GHz; smaller dust indices produce negative residuals at frequencies around 40 GHz. The low-resolution measurement at 102.5 GHz (smallest symbol) was excluded from the fit, as the anomalously high flux may include material outside the dust disc. The wind component is $0.27\, \nu_{GHz}^{0.23}$ mJy in the lower curve, and varies by +45% in the upper curve[20,35].

*(Middle:)* HD 97048. The dust is fitted with a $\nu^{3.25}$ spectrum, scaled to 70 mJy at 100 GHz. This fit demonstrates negligible AME at frequencies above 33 GHz; higher dust spectral indices are possible, yielding greater AME. A wind component with a $0.25\, \nu_{GHz}^{-0.1}$ mJy spectrum has been added, improving the fit at the limits set by the 2σ lowest-frequency data points.

*(Bottom:)* MWC 297. The dust component is fitted as a $\nu^2$ power-law spectrum, scaled to 127 mJy at 100 GHz. The fit assumes negligible AME at frequencies ≥72 GHz; relaxing this to higher frequency yields fits with more AME. The wind was fitted from our 5.5 and 8.8 GHz data with a $\nu^{0.28}$ power-law, while prior results at 4.9 and 8.4 GHz yielded a $\nu^{0.16}$ dependence. The two curves show these different cases when added to the dust signal (with the wind contributing 17-20% of the total signal at 100 GHz).

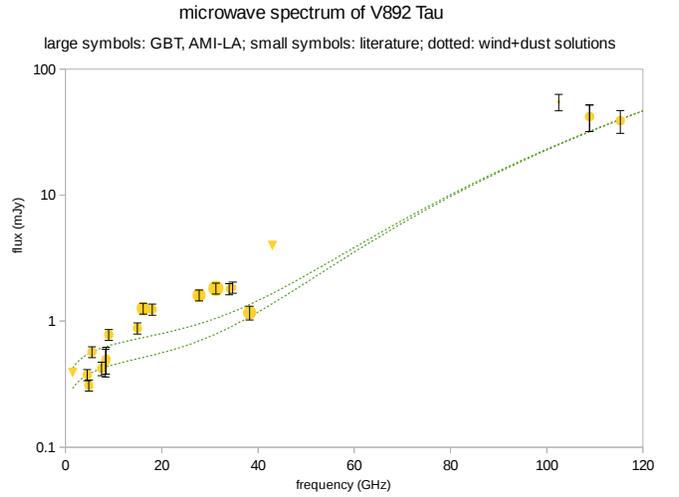

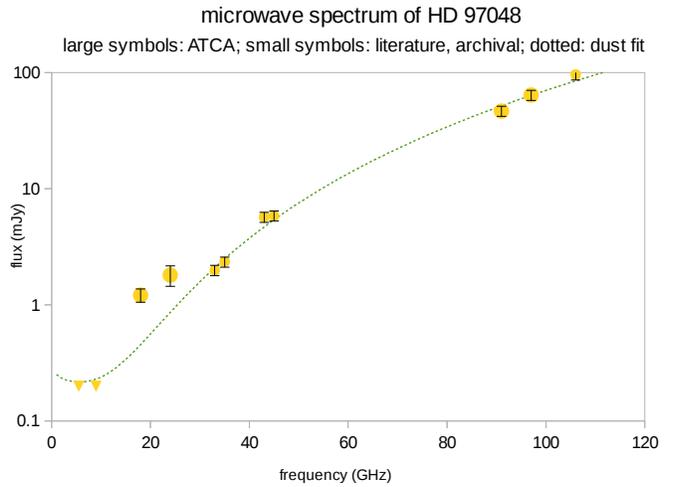

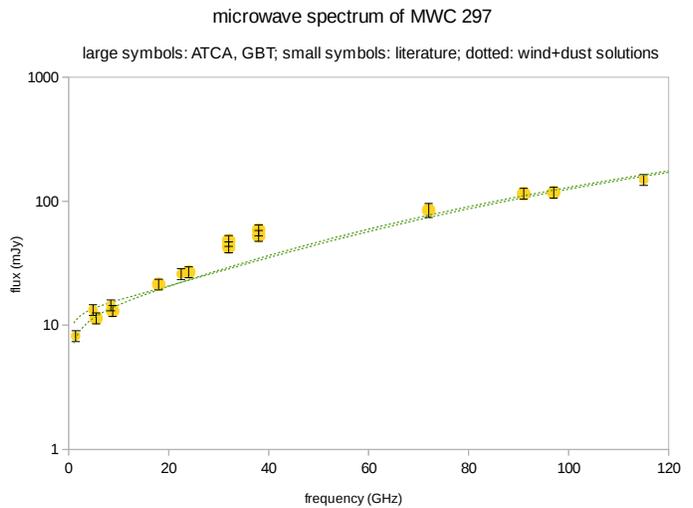

**Fig S3, cont.** Log-flux versus frequency for the systems where AME was not detected. Fits are illustrative, and not intended to account for higher-frequency dust measurements or time-variable wind emission discussed in the literature. New flux measurements are listed in Table S2.

*Sub-plots (a) to (d):* GBT sample, with additional literature fluxes (small symbols).
(a) AB Aur: $\nu^{3.02}$ dust spectrum scaled to 9.2 mJy at 100 GHz plus constant 0.12 mJy wind flux.
(b) IC 2087 IR: $\nu^{1.25}$ wind spectrum scaled to 32 µJy at 1 GHz; negligible dust.
(c) SR 21: fit to lowest wind limits (dot-dashed line) with $\nu^{3.4}$ dust of 5.0 mJy at 100 GHz, plus constant 25 µJy wind; alternative fit to wind detections (dotted line) and neglecting limits, i.e assuming wind levels may vary with time, with $\nu^{2.9}$ dust, 4.8 mJy at 100 GHz, 40 µJy wind.
(d) T Tau: $\nu^{4.0}$ dust, 30 µJy at 100 GHz, constant 11 mJy wind (but strongly time variable).

*Sub-plots (e) to (k):* ATCA sample, with supplementary ATCA fluxes[43] for HD 100546.
(e) CD-42 11721: $\nu^{0.7}$ wind spectrum scaled to 0.4 mJy at 1 GHz; negligible dust, using the detections only, as the region is confused by multiple sources, and no single-object fit is possible.
(f) HD 34282: $\nu^{2.0}$ dust, 4.8 mJy at 100 GHz, constant 50 µJy wind.
(g) HD 95881: maximised model fitted to limits, $\nu^{0.6}$ wind spectrum scaled to 40 µJy at 1 GHz.
(h) HD 100453: maximised model fitted to limits, $\nu^{0.8}$ wind spectrum scaled to 25 µJy at 1 GHz.
(i) HD 100546: $\nu^{2.7}$ dust spectrum scaled to 58 mJy at 100 GHz plus constant 0.5 mJy wind flux.
(j) HD 139614: $\nu^{4.0}$ dust spectrum, 0.36 mJy at 100 GHz, $\nu^{0.65}$ wind spectrum, 70 µJy at 1 GHz.
(k) HD 169142: $\nu^{4.0}$ dust spectrum, 10.5 mJy at 100 GHz, $\nu^{0.35}$ wind spectrum, 100 µJy at 1 GHz.

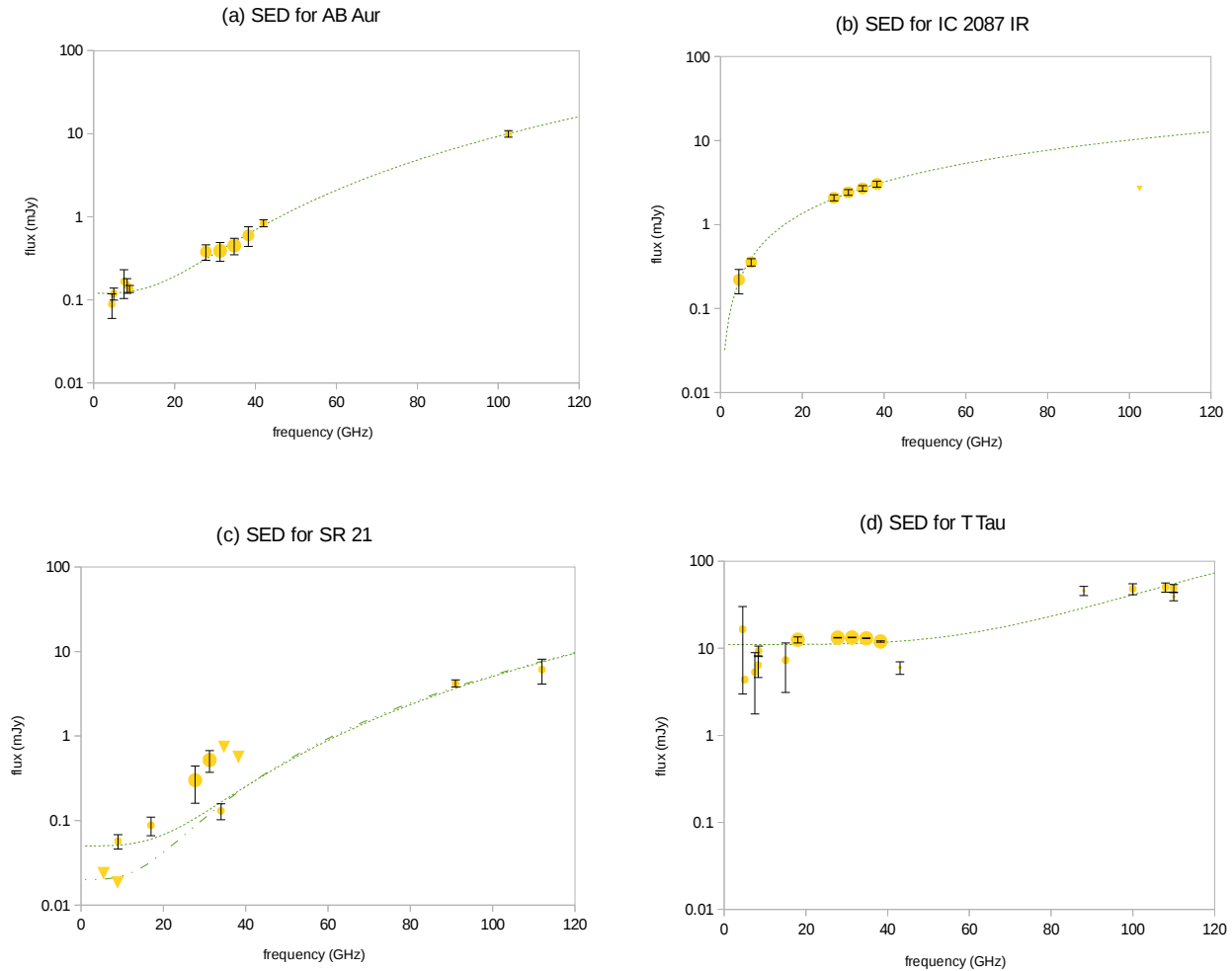

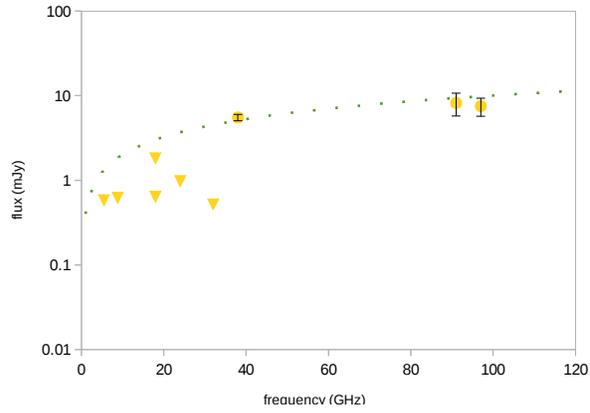
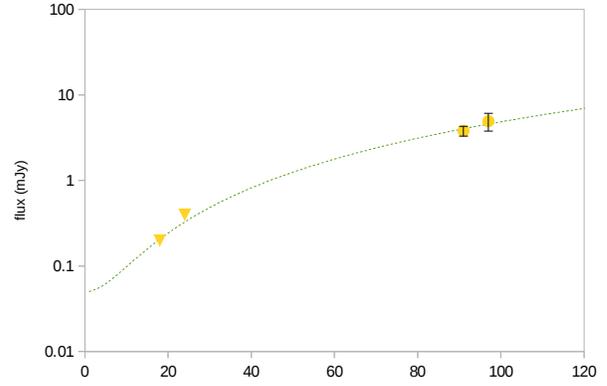
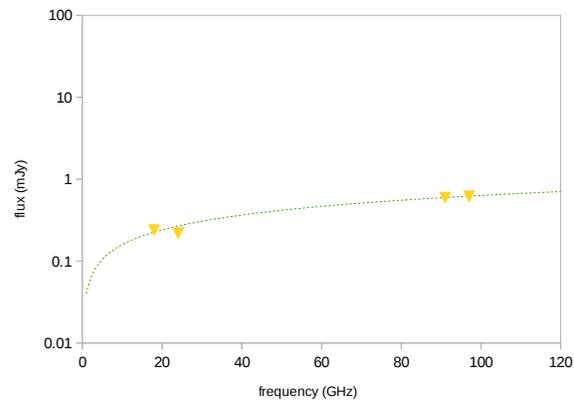
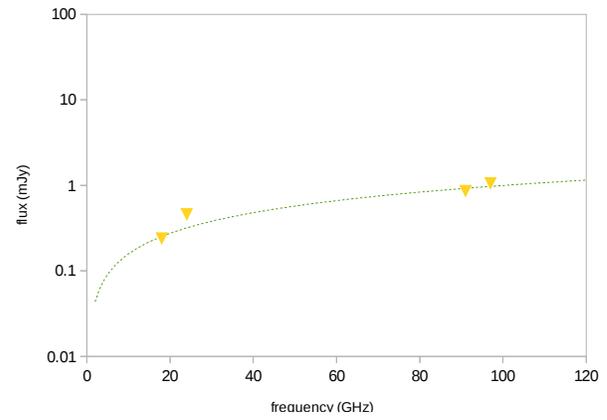
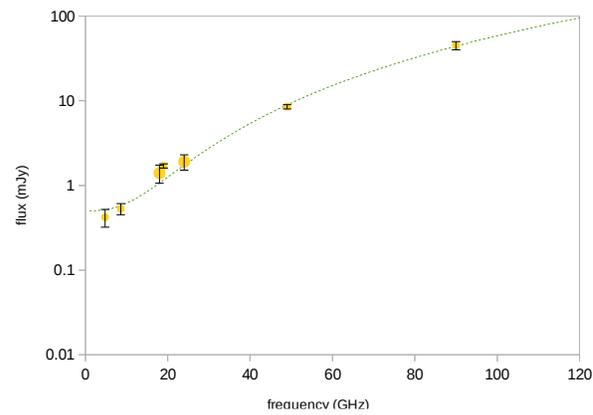
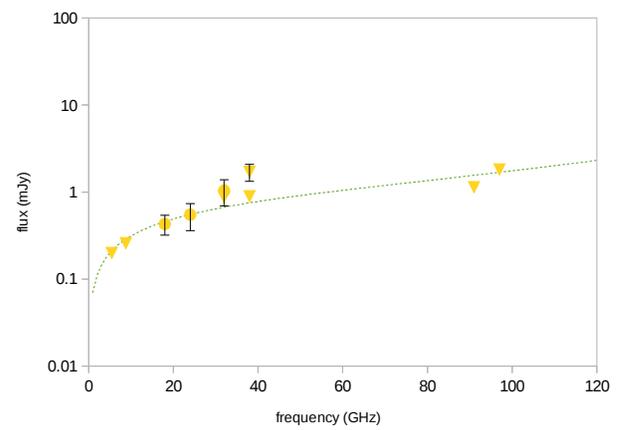
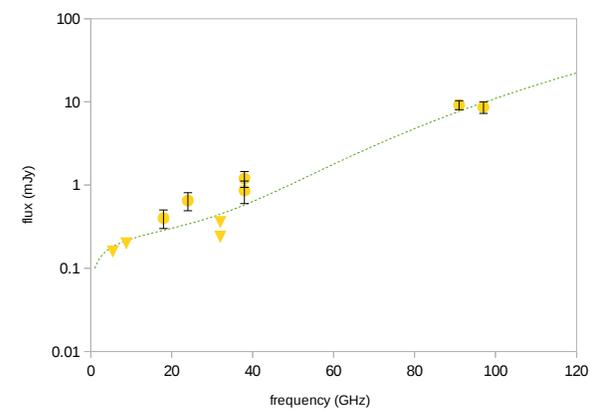

**Fig. S4**: *ISO-SWS* spectra towards the three diamond-hosting discs. For clarity, the V892 Tau data were binned over four 0.3 nm spectral channels, and 60 Jy was subtracted across the band for MWC 297. The nano-diamond peak wavelengths vary by ~±1 nm although this is near the limit of the spectral resolution; these shifts could indicate temperatures differences of ~100 K, within the ~800-1400 K range where hydrogenated nano-diamonds are stable[7,42].

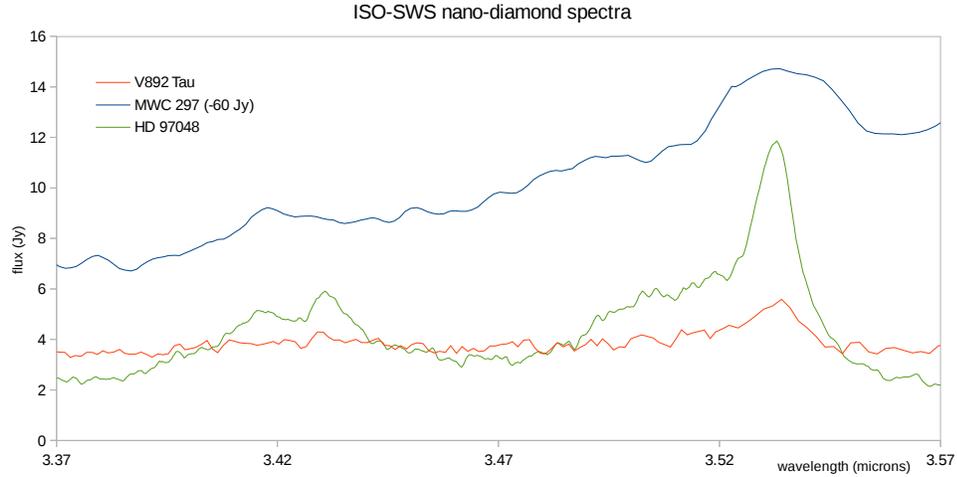

**Fig. S5**: Spectra from *Spitzer-IRS* (or *ISO-SWS* where not observed by *Spitzer*) showing the PAH features around 6.2 μm, for all the Herbig Ae stars observed within our sample (AME-detected objects shown with thick curves).

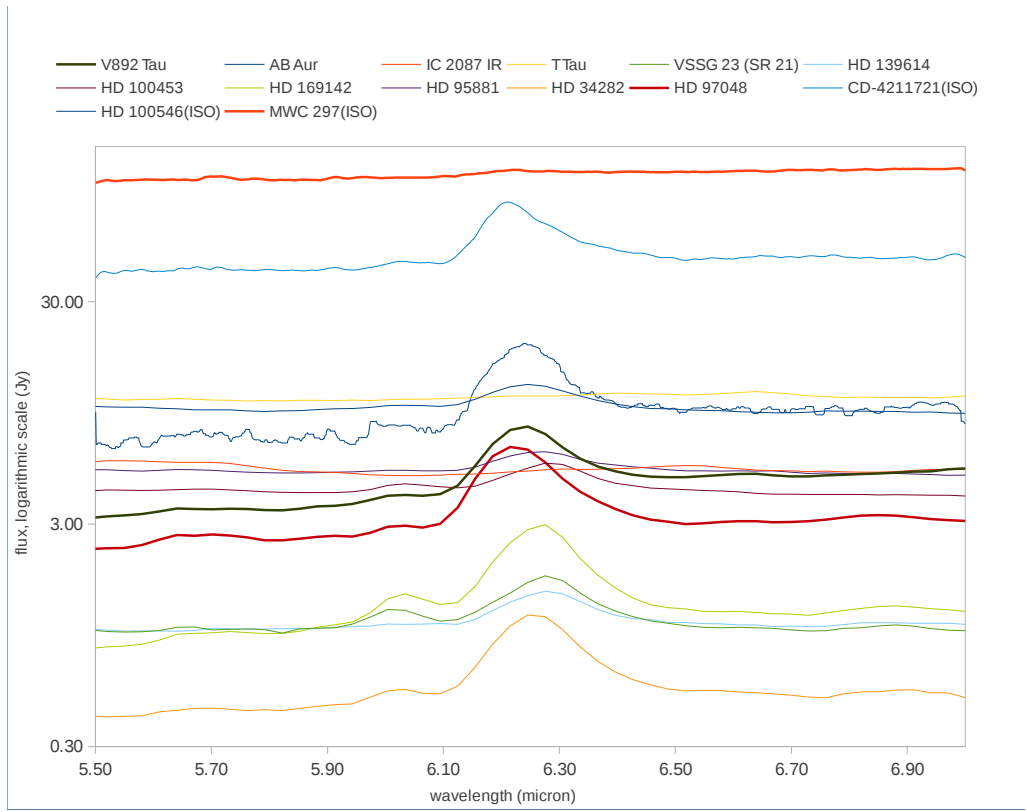

**Fig. S6.** Comparison of flux between the AME and nano-diamond 3.53 micron feature (left panel) and PAH 6.2 micron feature (right panel). The systems with detected AME are shown by the large blue circle symbols. In the comparison systems (shown with small yellow circles) AME upper limits are $< 1.5 \times 10^{-19}$ W m$^{-2}$ and the nano-diamond fluxes are $< 100 \times 10^{-16}$ W m$^{-2}$. There is no such distinction for the PAH signal, i.e. there are systems without AME whose PAH line flux is higher than the AME-systems V892 Tau and HD 97048.

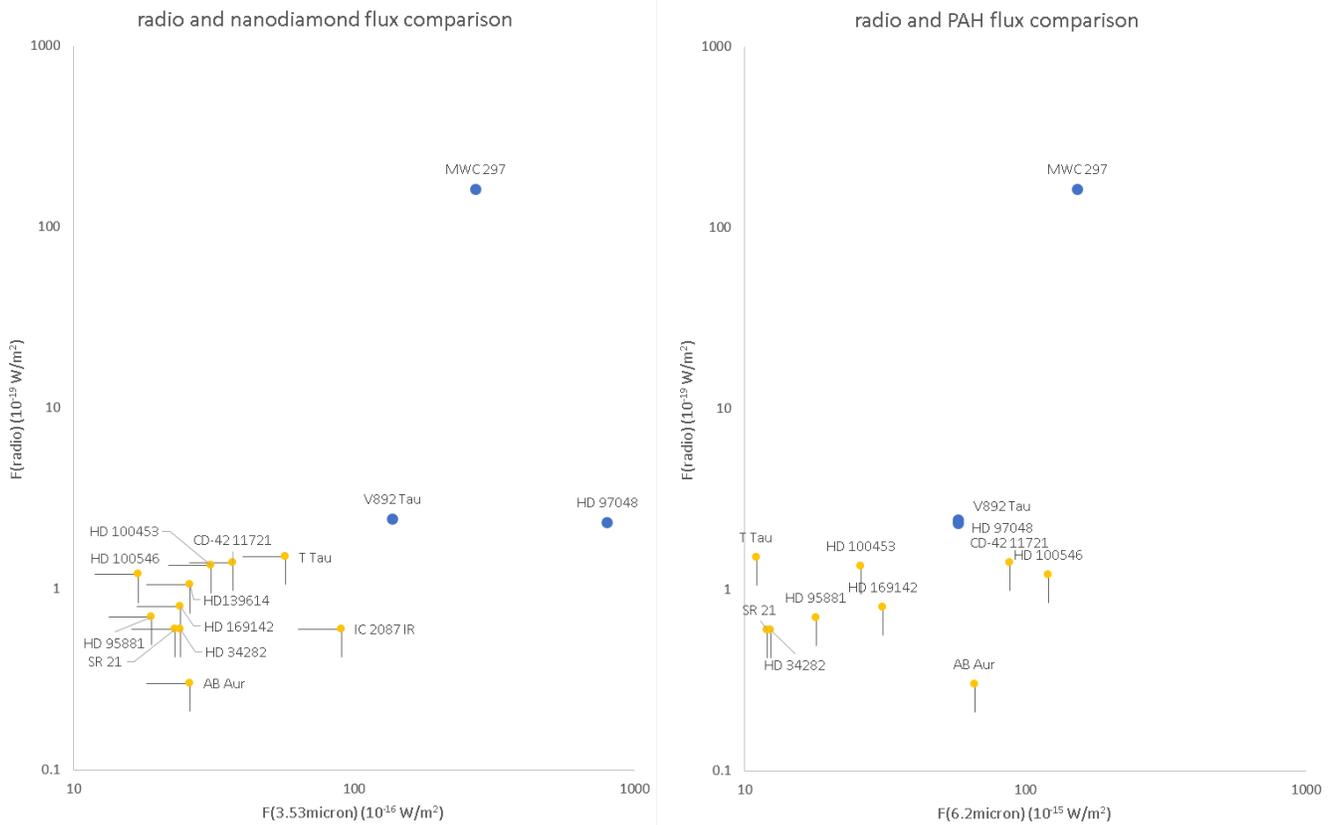

## AME Modelling

We fit for the AME carrier-particle parameters using two burn-in periods of 1000 steps each, followed by a production run, of the EMCEE EnsembleSampler, using 12 chains each of 10000 steps. We calculate the autocorrelation length, τ, using the ACOR package and discard the first 100τ samples before calculating the 16, 50 & 84% quantiles from the 1D posteriors to extract the expectation value of each parameter and its one sigma bounds. We use the raw likelihoods within the prior volume to extract the maximum likelihood parameter values. Each plot in Figures 1-3 shows maximum likelihood and expectation fits to the AME spectrum, along with 24 sample fits randomly drawn from the posterior. The parameter plots show the probability distributions of nano-diamond abundance and radius, and representative uncertainties are also listed in Table 1.

**Table S1.** Results of observations of the sources with AME, from interferometry with ATCA, AMI-LA and ALMA, plus photometry with GBT[44]. The flux errors quoted are from the rms noise in the measurements, with error in fitting a point source model added in quadrature where appropriate, but excluding calibration uncertainties (which are incorporated in errorbars in Fig. S3). Upper limits are 2-sigma, the same confidence level as adopted for the literature data in Fig. S3. Fluxes marked (p) or (s) indicate use of the primary or secondary calibrator.

| flux densities (mJy) | V892 Tau | MWC 297 | HD 97048 | HD 97048 (archival data) |
|---|---|---|---|---|
| F(5.5 GHz) (ATCA) | | 11.44 ± 0.24 | ≤ 0.2 | |
| F(8.8 GHz) (ATCA) | | 13.11 ± 0.24 | ≤ 0.2 | |
| F(16.1 GHz) (AMI-LA) | 1.26 ± 0.05 | | | |
| F(18 GHz) (ATCA) | | 21.40 ± 0.11 (s) | 1.21 ± 0.16 (s) | |
| F(24 GHz) (ATCA) | | 26.86 ± 0.10 (s) | 1.80 ± 0.36 (s) | |
| F(27.75 GHz) (GBT) | 1.61 ± 0.15 | | | |
| F(31.25 GHz) (GBT) | 1.82 ± 0.17 | | | |
| F(32 GHz) (ATCA) | | 42.71 ± 0.58 (s) <br> 48.11 ± 0.62 (p) | | |
| F(33 GHz) (ATCA) | | | | 2.0 ± 0.1 |
| F(34.75 GHz) (GBT) | 1.85 ± 0.17 | | | |
| F(35 GHz) (ATCA) | | | | 2.3 ± 0.1 |
| F(38 GHz) (ATCA) | | 52.8 ± 4.9 (s) <br> 58.6 ± 0.9 (p) | | |
| F(38.25 GHz) (GBT) | 1.17 ± 0.15 | | | |
| F(43 GHz) (ATCA) | | | | 5.7 ± 0.3 |
| F(45 GHz) (ATCA) | | | | 5.8 ± 0.3 |
| F(72 GHz) (GBT) | | 85 ± 11 | | |
| F(91 GHz) (ATCA) | | 115.4 ± 2.2 (p) | 46.5 ± 3.8 (p) | |
| F(97 GHz) (ATCA) | | 117.8 ± 6.5 (p) | 64.0 ± 6.0 (p) | |
| F(106 GHz) (ALMA) | | | | 96 ± 3 |

**Table S2.** New flux measurements for the sources without AME. $F_x$ indicates flux in mJy at x GHz; multiple flux values indicate different calibrators. Errors for ATCA include the rms noise and point-source model fit uncertainty added in quadrature. Limits are 2σ, as in Table S1.

| source | facility | flux data |
|---|---|---|
| AB Aur | GBT | $F_{27.75} = 0.38 \pm 0.08$, $F_{31.25} = 0.39 \pm 0.10$, $F_{34.75} = 0.45 \pm 0.10$, $F_{38.25} = 0.60 \pm 0.16$ |
| IC 2087 IR | GBT | $F_{27.75} = 2.08 \pm 0.17$, $F_{31.25} = 2.42 \pm 0.19$, $F_{34.75} = 2.70 \pm 0.20$, $F_{38.25} = 3.04 \pm 0.25$ |
| SR 21 | GBT | $F_{27.75} = 0.30 \pm 0.14$, $F_{31.25} = 0.52 \pm 0.15$, $F_{34.75} = 0.37 \pm 0.25$, $F_{38.25} = 0.18 \pm 0.19$ |
| T Tau | GBT | $F_{27.75} = 13.21 \pm 0.11$, $F_{31.25} = 13.26 \pm 0.11$, $F_{34.75} = 13.00 \pm 0.12$, $F_{38.25} = 11.95 \pm 0.21$ |
| CD-42 11721 | ATCA | $F_{5.5} \leq 0.58$, $F_{8.8} \leq 0.62$, $F_{18} \leq 0.64, 1.82$, $F_{24} \leq 0.98$, $F_{32} \leq 0.52$, $F_{38} = 5.52 \pm 0.47$, $F_{91} = 8.22 \pm 2.45$, $F_{97} = 7.53 \pm 1.82$ |
| HD 34282 | ATCA | $F_{18} \leq 0.20$, $F_{24} \leq 0.40$, $F_{91} = 3.81 \pm 0.50$, $F_{97} = 4.94 \pm 1.15$ |
| HD 95881 | ATCA | $F_{18} \leq 0.24$, $F_{24} \leq 0.22$, $F_{91} \leq 0.60$, $F_{97} \leq 0.62$ |
| HD 100453 | ATCA | $F_{18} \leq 0.24$, $F_{24} \leq 0.46$, $F_{91} \leq 0.86$, $F_{97} \leq 1.06$ |
| HD 100546 | ATCA | $F_{18} = 1.40 \pm 0.34$, $F_{24} = 1.91 \pm 0.39$ |
| HD 139614 | ATCA | $F_{5.5} \leq 0.20$, $F_{8.8} \leq 0.26$, $F_{18} = 0.43 \pm 0.11$, $F_{24} = 0.55 \pm 0.19$, $F_{32} = 1.04 \pm 0.35$, $\leq 0.88$, $F_{38} = 1.71 \pm 0.38$, $\leq 0.90$, $F_{91} \leq 1.14$, $F_{97} \leq 1.82$ |
| HD 169142 | ATCA | $F_{5.5} \leq 0.16$, $F_{8.8} \leq 0.20$, $F_{18} = 0.40 \pm 0.10$, $F_{24} = 0.65 \pm 0.16$, $F_{32} \leq 0.36$, $\leq 0.24$, $F_{38} = 1.20 \pm 0.26$, $0.86 \pm 0.26$, $F_{91} = 9.18 \pm 1.16$, $F_{97} = 8.65 \pm 1.38$ |

**Table S3.** Sample results and parameters (AME-systems in bold in the top row). The second row lists the AME flux in units of $10^{-19}$ W m$^{-2}$. AME limits were estimated from areas under a quasi-triangular spectrum, with a base ≤30 GHz wide, by a height of the greater of twice the maximum offset above the dust-plus-wind fit or twice the largest rms in the frequency range 18-38 GHz. The third row lists flux or 5σ limit (a conservative choice reflecting ground versus space discrepancies) for the 3.53 μm nano-diamond feature[16], in $10^{-16}$ W m$^{-2}$, with our derived ground, *ISO* estimates marked *, # respectively. The fourth row lists fluxes in the same unit estimated for the PAH 6.2 micron feature with *Spitzer*[6] or *ISO*[16]; our *Spitzer*-archive estimates are marked † and are uncertain for IC 2087 IR and T Tau where the feature may be a blend. The last three rows list literature estimates of distances (in pc); stellar luminosities (solar units) to facilitate comparison with $L_{radio}$, $L_{IR}$; and stellar effective temperatures (in K). L* estimates[12] have not been corrected for refinements in $T_{eff}$ values[47] nor do they take into account new parallaxes from *GAIA* (in data release 1); *GAIA* distance estimates are noted in brackets, in cases of major change only.

| | AB Aur | CD-42 11721 | HD 34282 | HD 95881 | **HD 97048** | HD 100453 | HD 100546 | HD 139614 | HD 169142 | IC 2087 IR | **MWC 297** | SR 21 | T Tau | **V892 Tau** |
|---|---|---|---|---|---|---|---|---|---|---|---|---|---|---|
| $F_{AME}$ | ≤ 0.3 | ≤ 1.4 | ≤ 0.6 | ≤ 0.7 | 2.3 | ≤ 1.4 | ≤ 1.2 | ≤ 1.1 | ≤ 0.8 | ≤ 0.6 | 160 | ≤ 0.6 | ≤ 1.5 | 2.4 |
| $F_{3.53}$ | ≤ 26 | ≤ 37 | ≤ 24 | ≤24 | 800 | ≤ 31 | ≤ 17 | ≤ 26 | ≤ 24 | ≤ 90* | 270 | ≤ 23* | ≤ 57# | 140 |
| $F_{6.2}$ | 660 | 880 | 120 | 180† | 580 | 260 | 1210 | 70 | 310 | ~90† | 1540 | 120† | ~110† | 580 |
| d | 139 | >136? (>1000) | ~350 | 170 (935) | 158 | 122 (243) | 97 | 142 | 145 | 140 | 250 | 120 | 148 | 142 |
| L* | 60 | ~1x10$^4$ | 14 | 27 | 40 | 14 | 24 | 9 | 11 | ~72 | ~2x10$^4$ | 6 | ~17 | 80 |
| $T_{eff}$ | 9800 | ~14000 | 9500 | 10000 | 10500 | 7250 | 9750 | 7750 | 7500 | ? | 24500 | 5950 | ? | 11200 |

**References for Additional Information:**